\documentclass[conference]{IEEEtran}
\IEEEoverridecommandlockouts
\usepackage{cite}
\usepackage{amsmath,amssymb,amsfonts}
\usepackage{algorithmic}
\usepackage{graphicx}
\usepackage{textcomp}
\usepackage{xcolor}
\usepackage{booktabs}
\usepackage{tabularx}
\usepackage{tabu}
\usepackage{array}
\usepackage{siunitx}

\def\BibTeX{{\rm B\kern-.05em{\sc i\kern-.025em b}\kern-.08em
    T\kern-.1667em\lower.7ex\hbox{E}\kern-.125emX}}
\begin{document}

\title{SSM-RDU: A Reconfigurable Dataflow Unit for Long-Sequence State-Space Models}

\author{\IEEEauthorblockN{Sho Ko}
\IEEEauthorblockA{\textit{I Machines, Inc} \\
Santa Clara, CA, USA \\
kosho2013@hotmail.com}
\and
\IEEEauthorblockN{Kunle Olukotun}
\IEEEauthorblockA{\textit{Department of Electrical Engineering} \\
\textit{Stanford University}\\
Stanford, CA, USA \\
kunle@stanford.edu}
}

\maketitle

\begin{abstract}
Long-sequence state-space models (SSMs) such as Hyena and Mamba replace the quadratic complexity of self-attention with more efficient FFT and scan operations.
However, modern accelerators like GPUs are poorly suited to these non-GEMM workloads due to rigid execution models and specialization for dense matrix operations.
This paper proposes architectural extensions to a baseline Reconfigurable Dataflow Unit (RDU) that efficiently support FFT-based and scan-based SSMs.
By introducing lightweight interconnect enhancements within compute tiles, the extended RDU enables spatial mapping of FFT and scan dataflows with less than 1\% area and power overhead.
The resulting architecture achieves a $5.95\times$ speedup over the NVIDIA A100 GPU and a $1.95\times$ speedup over the baseline RDU for Hyena, and a $2.12\times$ and $1.75\times$ speedup over the A100 GPU and baseline RDU, respectively, for Mamba.
\end{abstract}

\section{Introduction}

Self-attention-based large language models (LLMs)~\cite{vaswani2017attention} have revolutionized the field of machine learning but they suffer from the quadratic computational complexity $O(N^2)$ relative to the sequence length, making them impractical for long-sequence modeling.
However, long-sequence modeling is crucial for application domains requiring high-resolution temporal understanding such as genomics and bio-informatics~\cite{nguyen2023hyenadna}, audio processing~\cite{goel2022s}, video understanding~\cite{li2024videomamba}, financial forecasting~\cite{rigatos2017state}, and weather modeling~\cite{adedotun2020modelling}.
State-space models (SSMs) such as Hyena~\cite{nguyen2023hyenadna, poli2023hyena, massaroli2023laughing} and Mamba~\cite{gumamba, dao2024transformers} are promising solutions for long-sequence modeling since they replace the quadratic attention computation with sub-quadratic log-linear $O(Nlog_{2}(N))$ computation patterns such as FFT and scan.
SSMs can scale up to a sequence length of one million.
Modern GPUs are ill-suited for SSM workloads due to two key constraints.
First, their computational throughput is dominated by tensor cores, which contribute up to $\sim80\%$ of their peak theoretical FLOPs~\cite{nvidiaH100datasheet}.
While tensor cores excel at dense matrix multiplication (GEMM), they are inherently inefficient for non-GEMM kernels such as FFT kernels in Hyena and scan kernels in Mamba, which involve irregular memory access patterns, data-dependent computation, and low arithmetic intensity.
Forcing non-GEMM kernels on tensor cores in GPUs results in sub-optimal algorithm~\cite{fuflashfftconv} or significant software change~\cite{dakkak2019accelerating}.
Second, GPUs suffer from limited kernel fusion capabilities due to their rigid execution model, which processes kernel-by-kernel sequentially.
This forces intermediate results to be staged in off-chip memory, incurring significant latency and energy overheads~\cite{prabhakar2021sambanova}.
To address these limitations, people have proposed a Reconfigurable Dataflow Unit (RDU) architecture optimized for diverse computational patterns in both academia~\cite{prabhakar2017plasticine} and industry~\cite{prabhakar2021sambanova, prabhakar2024sambanova}.
Unlike GPUs, which centralize compute resources and memory hierarchies, the RDU employs a distributed pattern compute units (PCUs) and pattern memory units (PMUs) coupled with programmable network-on-chip (NoC) switches.
We start with a baseline architecture of a RDU, which can dynamically reconfigured to support different computation patterns such as SIMD and systolic, and extend it to support FFT and scan operations.
With such design, two critical advantages are enabled: kernel fusion and adaptive parallelism.
For kernel Fusion, multiple kernels reside on the chip simultaneously and data are streamed and pipelined across kernels on-chip, eliminating off-chip memory bottlenecks and enabling fusion of multiple kernels.
For adaptive parallelism, RDU compute tiles can be reconfigured at runtime to match the dataflow requirements of FFT (butterfly operations) and scan (prefix-sum trees), maximizing parallelism while minimizing control overheads.
We summarize the key contributions of this work as follows:
\begin{itemize}
    \item We present an in-depth analysis of the RDU architecture, along with a detailed study of Hyena SSM and Mamba SSM algorithms.

    \item We propose an extension to the baseline RDU to support FFT operations, enabling an efficient mapping of the Hyena SSM onto the RDU.
    This implementation achieves a $1.95\times$ speedup over the baseline RDU and a $2.0\times$ to $5.95\times$ speedup over a baseline GPU, while incurring less than 1\% area and power overhead.

    \item We propose a second extension to baseline RDU to support scan operations, facilitating the mapping of the Mamba SSM onto the RDU.
    This approach achieves a $1.75\times$ speedup over the baseline RDU and a $2.12\times$ speedup over a baseline GPU, also with less than 1\% area and power overhead.
\end{itemize}

\section{Background}

\subsection{Reconfigurable Dataflow Architecture}

\begin{figure}[t!]
  \centering
  \includegraphics[width=7cm]{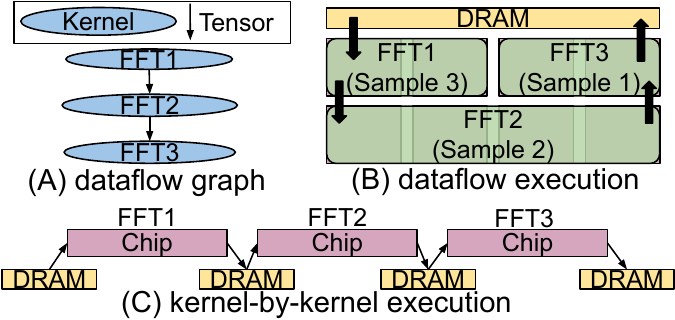}
  \vspace{-10pt}
  \caption{(A) The workload dataflow graph in which vertices represent computation kernels and edges represent tensors.
  (B) Dataflow execution: multiple kernels are fused on-chip and data is pipelined through the kernels in a streaming fashion.
  (C) Kernel-by-kernel execution: kernels are executed sequentially with frequent DRAM accesses between kernels.}
  \vspace{-10pt}
  \label{dataflow}
\end{figure}

\begin{figure}[t!]
  \centering
  \includegraphics[width=\linewidth]{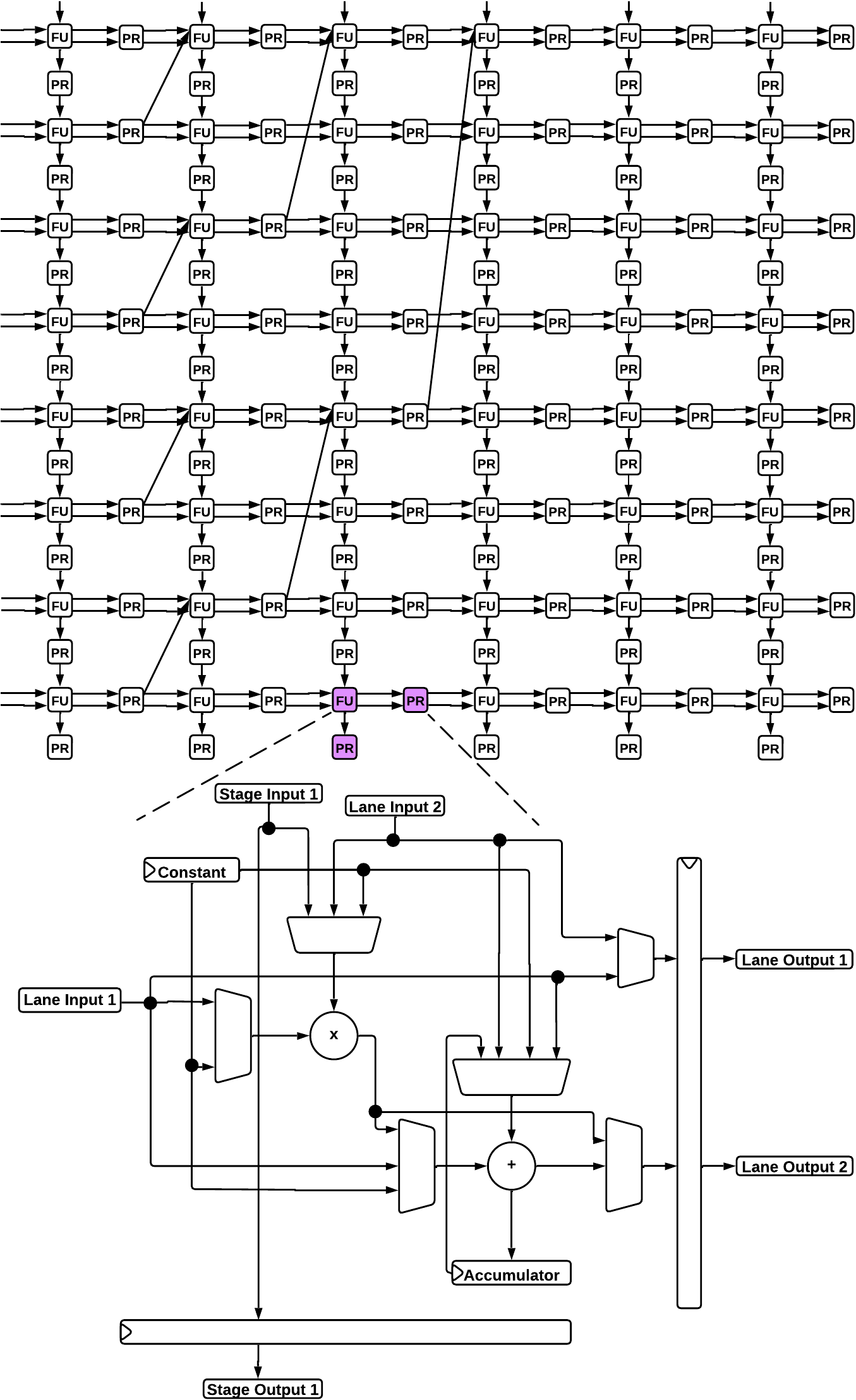}
  \vspace{-10pt}
  \caption{The RDU chip is composed of a grid of PCUs and PMUs.
  Each PCU features a pipelined SIMD architecture with multiple lanes and multiple pipeline stages.
  It supports three execution modes: element-wise, systolic, and reduction.
  Within each functional unit (FU), there are four input sources, along with dedicated add and multiply operations. The FU can be configured to perform a multiply-and-accumulate (MAC) operation, or a scalar addition or multiplication between any two of the four input sources.}
  \label{rdu}
  \vspace{-10pt}
\end{figure}

When executing a workload dataflow graph shown like the one in Figure~\ref{dataflow}A, different accelerators can be classified into two categories of execution models: kernel-by-kernel and dataflow.
Kernel-by-kernel execution is typically done in instruction-based processor architectures such as NVIDIA's graphics processing unit (GPU)~\cite{choquette2020nvidia, choquette2022nvidia}.
Kernel-by-kernel execution loads the data from memory to on-chip, executes the kernel, and then stores the results back to memory, as shown in Figure~\ref{dataflow}C.
This incurs more memory traffic and results in compute under-utilization.
Dataflow execution is typically done in spatial architectures like SambaNova's reconfigurable dataflow unit (RDU)~\cite{prabhakar2021sambanova, prabhakar2024sambanova}.
Dataflow execution is capable of mapping multiple kernels spatially, fusing them on-chip, and pipelining input data through the kernels in a streaming fashion, as shown in Figure~\ref{dataflow}B.
This spatial computing paradigm results in less memory traffic and improves compute utilization.
Zooming into the micro-architecture of the RDU, it comprises a grid of distributed compute tiles (PCUs) and memory tiles (PMUs).
Each PCU features a pipelined SIMD architecture with multiple lanes and stages: Figure~\ref{rdu} depicts an instance with 8 lanes and 6 pipeline stages.
The PCU supports three primary operation modes: element-wise, systolic, and reduction.
In the element-wise mode, data flows unidirectionally from left to right.
In the systolic mode, data flows both horizontally (left to right) and vertically (top to bottom), enabling efficient matrix-like computations.
In the reduction mode, data flows from left to right, leveraging an inter-stage reduction tree interconnect to aggregate partial results efficiently.
Each Functional Unit (FU) has a total of four input sources: two from the lane dimension, one from the stage dimension, and one constant input.
The FU supports three core operations: scalar multiplication, scalar addition, and multiply-and-accumulate (MAC).
The MAC operation is employed in the systolic mode configuration of the PCU, where it performs partial product accumulation and propagates results to neighboring FUs.
The scalar addition and multiplication operations enable flexible computation between any two of the four available inputs.
These scalar operations are primarily used in the element-wise and reduction modes of the PCU configuration.

\subsection{State-Space Models}

\begin{figure}[t!]
  \centering
  \includegraphics[width=\linewidth]{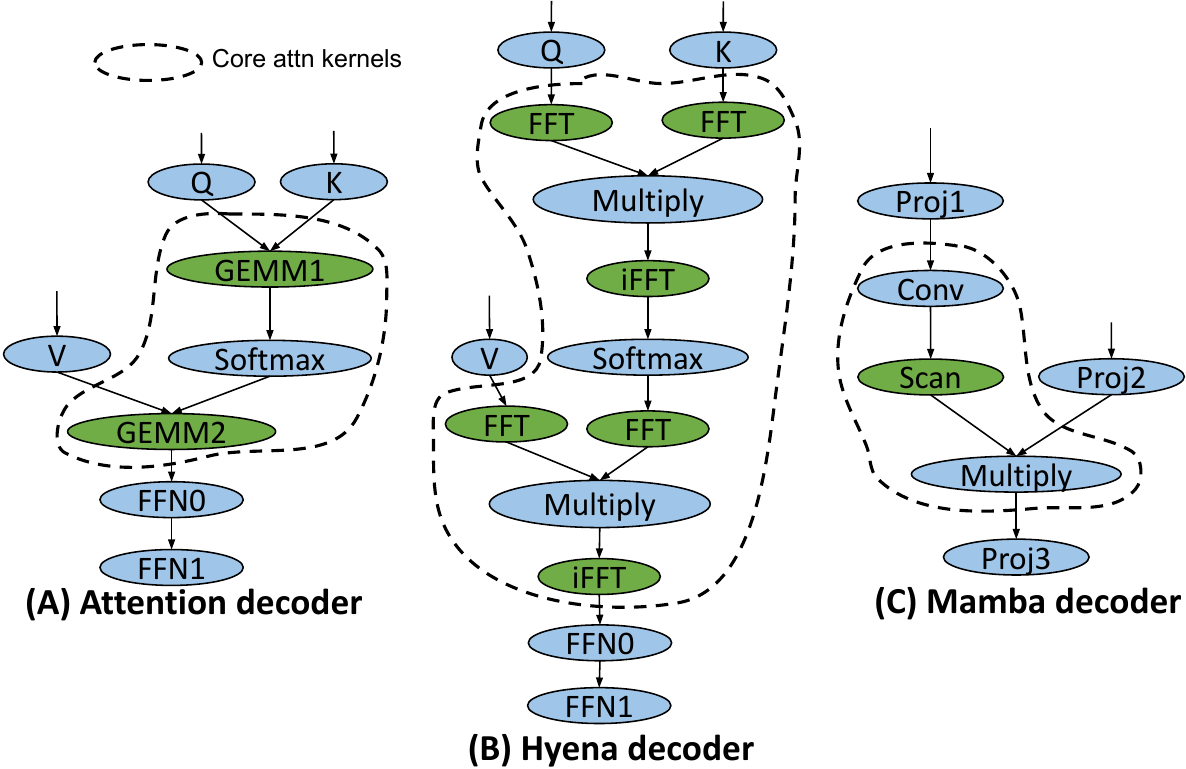}
  \vspace{-10pt}
  \caption{(A) Dataflow graph of an attention decoder, based on GEMM operations.
  (B) Dataflow graph of a Hyena decoder, based on FFT operations.
  (C) Dataflow graph of a Mamba decoder, based on scan operations.}
  \vspace{-10pt}
  \label{ssm}
\end{figure}

Figure~\ref{ssm} illustrates the workload dataflow graphs of an attention decoder layer, a Hyena decoder layer, and a Mamba decoder layer.
The attention decoder is structured around a quadratic-time GEMM kernel applied to the input QKV matrices.
The Hyena decoder adopts the same structural template as the attention decoder but replaces the GEMM kernel with an FFT-based convolution kernel~\cite{nguyen2023hyenadna, poli2023hyena, massaroli2023laughing}.
Specifically, each GEMM is replaced by three FFT operations: two forward FFTs to transform the input matrices from the time domain to the frequency domain, and one inverse FFT (iFFT) to convert the output back to the time domain.
In contrast, the Mamba decoder follows a slightly different architecture.
It is a linear time-invariant (LTI) model that evolves hidden states across the sequence~\cite{gumamba, dao2024transformers}.
Its core operation is a scan, which applies a recurrent state update across time steps in a sequential manner.

\subsection{Performance Model: DFModel}

% \begin{figure}[t!]
%   \centering
%   \includegraphics[width=\linewidth]{pdf/dfmodel.pdf}
%   \vspace{-10pt}
%   \caption{DFModel takes a workload and a system configuration as inputs, performs a multi-level optimization process to identify the optimal dataflow mapping, and estimates the corresponding performance.}
%   \vspace{-10pt}
%   \label{dfmodel}
% \end{figure}

We use DFModel~\cite{ko2024dfmodel} as the performance modeling framework to evaluate system performance in this work.
DFModel takes a workload and a system configuration as inputs, performs a multi-level optimization process to identify the optimal dataflow mapping, and estimates the corresponding performance.
DFModel supports the modeling of various hardware architectures, including RDUs and GPUs, as well as diverse memory technologies such as DDR and HBM.
It targets a wide range of workloads, including ML models such as LLMs and deep learning recommendation models (DLRMs)~\cite{mudigere2022software}, in addition to HPC workloads like FFTs and High Performance LINPACK (HPL)~\cite{kim2022snuhpl}.

\section{Map Hyena to RDU}
\label{hyena}

\begin{figure}[t!]
  \centering
  \includegraphics[width=\linewidth]{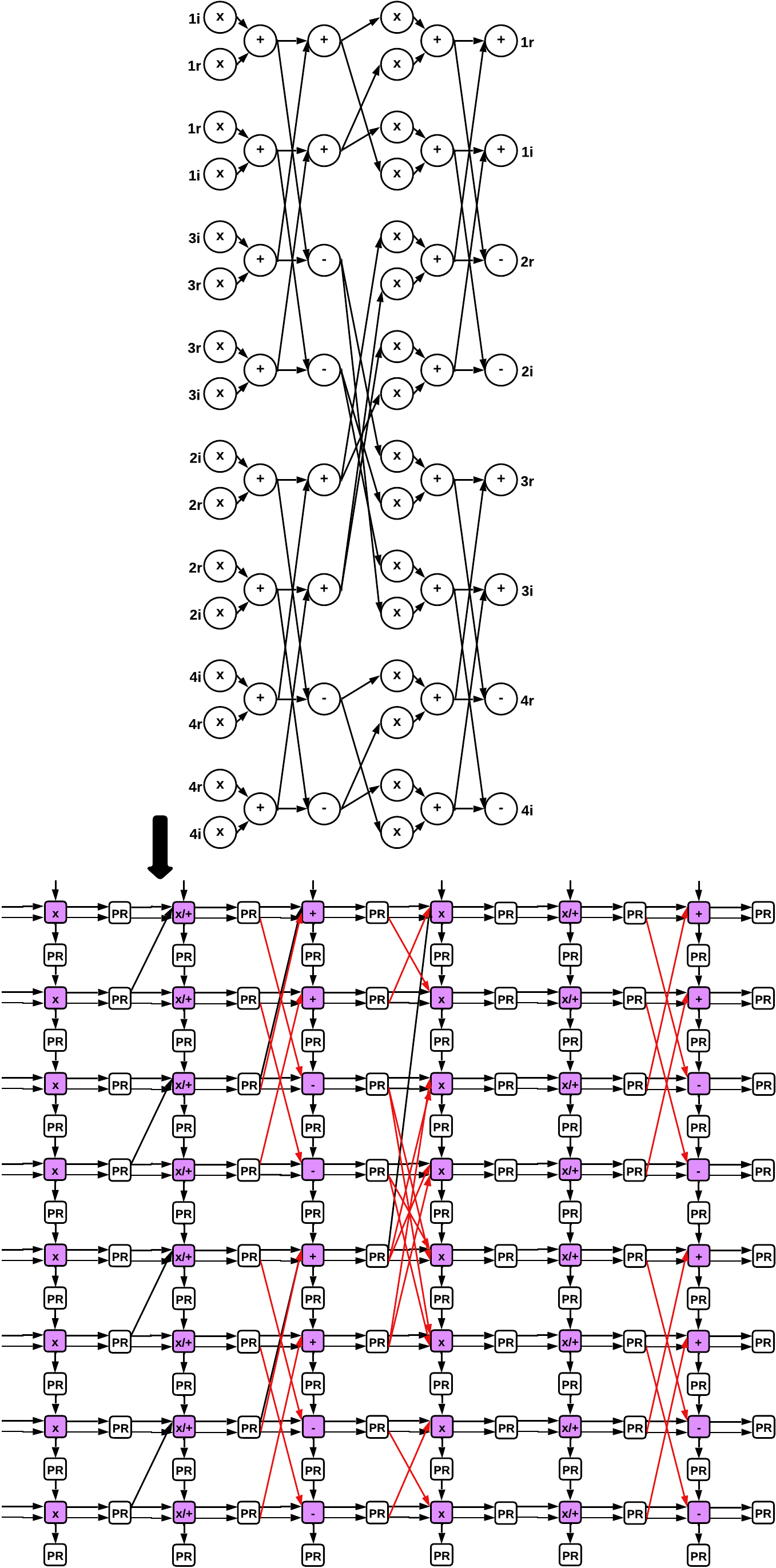}
  \vspace{-10pt}
  \caption{The figure illustrates the dataflow graph of a 4-point FFT alongside the proposed architectural extension to the PCU for supporting FFT execution.
  This extension incorporates butterfly interconnects between pipeline stages, enabling the FFT dataflow to be efficiently and directly mapped onto the PCU structure.}
  \vspace{-10pt}
  \label{fft_pcu}
\end{figure}

\begin{figure}[t!]
  \centering
  \includegraphics[width=\linewidth]{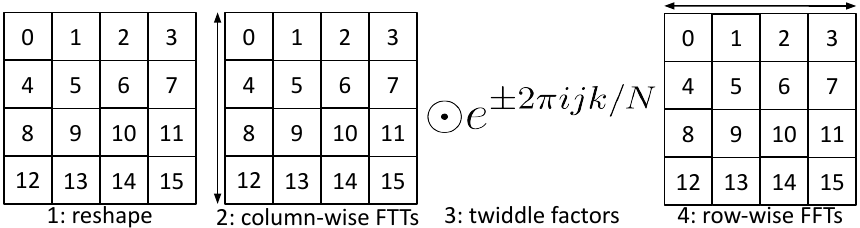}
  \vspace{-10pt}
  \caption{Bailey’s FFT algorithm consists of four main steps.
  First, the one-dimensional input sequence is reshaped into a two-dimensional matrix.
  Second, FFTs are independently computed along each column of the matrix.
  Third, element-wise multiplications are performed using the appropriate twiddle factors.
  Finally, FFTs are independently computed along each row.}
  \vspace{-10pt}
  \label{fft}
\end{figure}

In this section, we provide a detailed analysis of various FFT algorithm variants and their associated trade-offs.
Building upon this foundation, we propose a PCU architecture extension which can support FFTs efficiently.
Furthermore, we design an experiment to empirically evaluate and demonstrate the performance advantages of the proposed FFT-mode PCU.

\subsection{FFT Algorithms}

The Hyena decoder uses FFTs as its core computation kernel~\cite{nguyen2023hyenadna, poli2023hyena, massaroli2023laughing}.
The standard Cooley-Tukey FFT algorithm~\cite{cooley1965algorithm} is inefficient for modern hardware accelerators for several reasons~\cite{bailey1988high, fuflashfftconv}.
First, the variable-distance butterfly connections inherent to the algorithm hinder efficient vectorization.
Second, FFTs are not naturally expressed as dense matrix multiplication, which is the core computational primitive optimized in contemporary accelerator architectures.
Bailey's FFT algorithm~\cite{bailey1990ffts} is a specialized case of the Cooley–Tukey FFT algorithm, designed for enhanced efficiency on hardware architectures equipped with vector processing capabilities.
The algorithm partitions an $L$-point sequence into $\frac{L}{R}$ segments, each of length $R$.
As illustrated in Figure~\ref{fft}, Bailey's FFT comprises a four-step procedure. First, reshape the one-dimensional input sequence into a two-dimensional matrix.
Second, independently compute FFTs along each column.
Third, apply element-wise multiplications with the appropriate twiddle factors.
Fourth, independently compute FFTs along each row.
This structured approach optimizes data locality and vectorization, thereby improving performance on vector processors.
Bailey's method effectively addresses the challenges associated with non-unit stride memory accesses found in traditional FFT algorithms, leading to significant performance gains on vector supercomputers.
For each tile of length $R$ in Bailey's FFT algorithm, the value of $R$ is typically chosen to be 16 or 32 to match the width of vector lanes on modern architectures.
Depending on the method used to compute the $R$-point FFTs, Bailey's algorithm has two notable variants:
\begin{itemize}
    \item \textbf{Vector FFT:} Each $R$-point FFT is computed using an $O(Rlog_{2}(R))$ Cooley-Tukey FFT.
    The total number of floating-point operations (FLOP) for an $L$-point input sequence is $O(Llog_{2}(L))$, which is asymptotically optimal.
    However, this approach requires specialized hardware units to efficiently perform the $R$-point FFTs, such as a pipelined SIMD processing unit with butterfly interconnects (as proposed later in this section).

    \item \textbf{GEMM FFT:} Each $R$-point FFT is computed using a naive $O(R^2)$ Discrete Fourier Transform (DFT), resulting in a total FLOP count of $O(RLlog_{R}(L))$.
    Due to the DFT operations, the computational complexity is suboptimal ($\sim6.4\times$ more FLOP for $R=32$).
    However, this method is well-suited for acceleration using GEMM units, such as tensor cores on NVIDIA GPUs.
    Prior works~\cite{fuflashfftconv, li2021tcfft}, have demonstrated the feasibility of GEMM-based FFT implementations on GPU architectures.
\end{itemize}

\begin{figure}[t!]
  \centering
  \includegraphics[width=\linewidth]{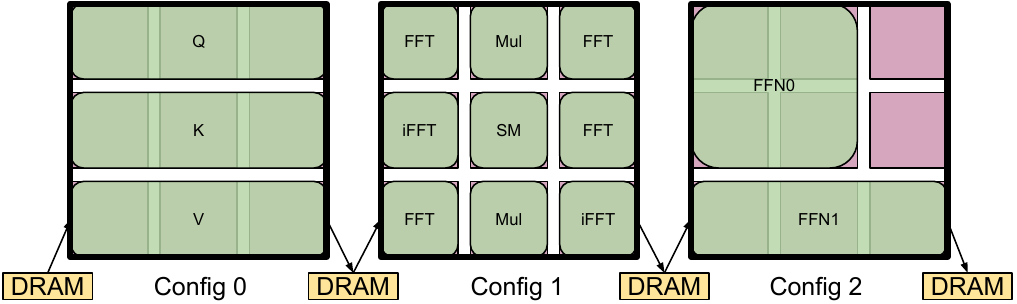}
  \vspace{-10pt}
  \caption{DFModel~\cite{ko2024dfmodel} provides a dataflow mapping of the Hyena decoder's dataflow graph, shown in Figure~\ref{ssm}, onto an RDU chip. It determines the optimal mapping that maximizes overall throughput while satisfying all on-chip compute and memory resource constraints.}
  \vspace{-10pt}
  \label{dataflow_mapping}
\end{figure}

\subsection{PCU FFT Mode}

For the baseline RDU-PCU architecture shown in Figure~\ref{rdu},
GEMM FFT executes efficiently, as it primarily consists of matrix multiplications that can fully leverage the PCU’s systolic execution mode.
In contrast, Vector FFT performs poorly on the baseline architecture due to the absence of cross-lane butterfly interconnects between pipeline stages.
The existing interconnects, which are designed for the reduction tree in the reduction mode, are insufficient to support the data movement patterns required by Vector FFT.
As a result, mapping Vector FFT onto the baseline PCU restricts execution to only the first stage of the pipeline, leading to low resource utilization and significantly degraded performance.
To enable efficient execution of Vector FFTs on PCUs, we propose an architectural extension illustrated in Figure~\ref{fft_pcu}.
This extension introduces butterfly interconnects between pipeline stages, allowing the FFT dataflow to be effectively mapped onto the PCU structure.
Figure~\ref{fft_pcu} depicts the dataflow graph of a 4-point FFT and its spatial mapping onto an $8\times6$ PCU.
In this mapping, the vertices, which represent arithmetic operations such as additions and multiplications, are assigned to functional units (FUs) within the PCU; the edges, which represent data dependencies, are realized through the interconnects.
The entire dataflow graph can be spatially unrolled across the PCU fabric, akin to an ASIC-style implementation.
This spatial mapping enables higher performance and improved energy efficiency compared to instruction-driven architectures such as CPUs and GPUs~\cite{tang2022high}.
To map the entire dataflow graph of the Hyena decoder in Figure~\ref{ssm} onto the RDU chip, as illustrated in Figure~\ref{dataflow}B, it is essential to optimally allocate resources to each kernel in the graph.
This ensures all on-chip compute and memory constraints are satisfied, resulting in a balanced pipeline and maximum overall throughput. 
Figure~\ref{dataflow_mapping} shows an example of such a dataflow mapping.
When the dataflow graph exceeds the capacity of a single RDU configuration, DFModel automatically partitions the graph into multiple subgraphs and maps each to a separate configuration of the RDU.

\subsection{Experimental Results}

\begin{table}[b!]
    \centering
    \vspace{-10pt}
    \caption{RDU architectural specification.}
    \begin{tabu}{p{3.5cm}p{3.5cm}}
    \rowfont{\bfseries} Specification & Value \\
    \midrule
    Compute & 520 PCUs, 32x12 each \\
    On-chip SRAM & 520 PMUs, 1.5 MB each \\
    Clock frequency & 1.6GHz, 640TFLOPS FP16 \\
    Off-chip DRAM & 8TB/s, HBM3e\\
    \end{tabu}
    \vspace{-10pt}
    \label{spec}
\end{table}

\begin{figure}[t!]
  \centering
  \includegraphics[width=\linewidth]{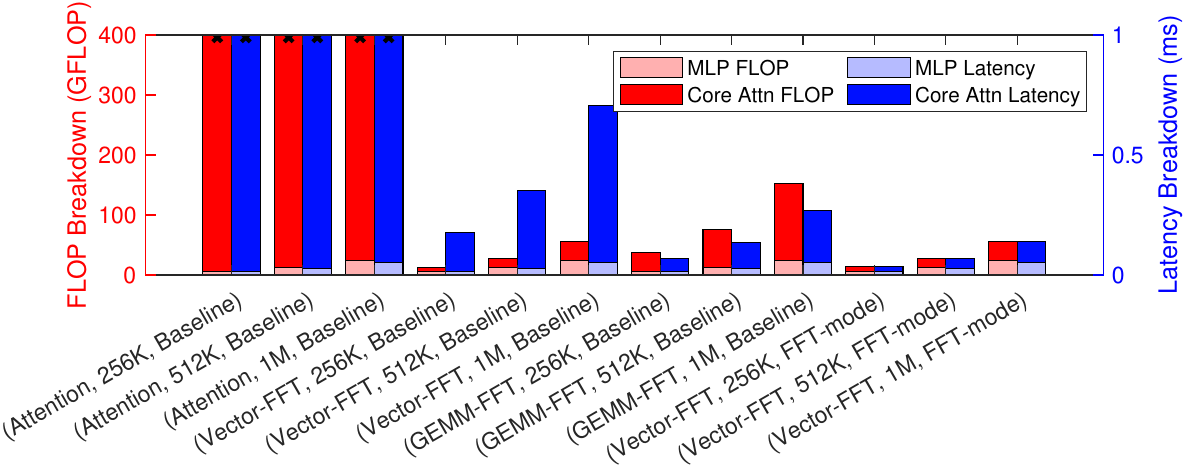}
  \vspace{-10pt}
  \caption{The figure compares four designs: (1) the attention decoder on the baseline RDU, (2) the Vector-FFT Hyena decoder on the baseline RDU, (3) the GEMM-FFT Hyena decoder on the baseline RDU, and (4) the Vector-FFT Hyena decoder on the FFT-mode RDU.
  Design (1) exhibits the highest latency.
  Design (2) achieves a $217.74\times$ speedup over Design (1).
  Design (3) design achieves a $2.61\times$ speedup over Design (2).
  Design (4) achieves a $1.95\times$ speedup over Design (3).}
  \vspace{-10pt}
  \label{hyena_time_flop}
\end{figure}

In our experiments, we evaluate four designs: (1) the attention decoder on the baseline RDU, (2) the Vector-FFT Hyena decoder on the baseline RDU, (3) the GEMM-FFT Hyena decoder on the baseline RDU, and (4) the Vector-FFT Hyena decoder on the FFT-mode RDU.
All decoders are configured with a hidden dimension of 32.
We sweep across three sequence lengths: 256K, 512K, and 1M. Using DFModel~\cite{ko2024dfmodel}, we map the different decoders to an RDU with architectural specifications described in Table~\ref{spec}.
Figure~\ref{hyena_time_flop} presents the FLOP count and latency breakdown for the four designs.
Several key observations can be drawn from the figure:
\begin{itemize}
    \item The attention decoder incurs a high FLOP count due to its quadratic attention mechanism, resulting in significant latency.
    
    \item The Vector-FFT Hyena decoder achieves the lowest theoretical FLOP count.
    However, the baseline RDU suffers from underutilization due to the absence of butterfly interconnects in its PCUs. Despite this, it attains a $217.74\times$ speedup over the attention decoder across various sequence lengths.
    
    \item The GEMM-FFT Hyena decoder exhibits a higher FLOP count, which is approximately $4.19\times$ greater than the Vector-FFT variant, due to the less efficient arithmetic structure.
    Nevertheless, it enables more efficient utilization of the baseline RDU via systolic mode execution of GEMM kernels, leading to a $2.61\times$ speedup over the Vector-FFT Hyena decoder on the same hardware.
    
    \item The Vector-FFT Hyena decoder deployed on the FFT-mode RDU benefits from both optimal FLOP count and high hardware utilization, enabled by enhanced PCU connectivity.
    As a result, it achieves a $1.95\times$ speedup over the GEMM-FFT Hyena decoder on the baseline RDU across different sequence lengths.
\end{itemize}

\begin{table}[t!]
    \centering
    \vspace{-10pt}
    \caption{Architectural specifications of three accelerators.}
    \begin{tabu}{p{3cm}p{1.3cm}p{1.3cm}p{1.3cm}}
    \rowfont{\bfseries}  & GPU & VGA & FFT RDU \\
    \midrule
    GEMM FP16 TFLOPS & 311.87 &	655.36 &	638.98 \\
    FFT FP16 TFLOPS & 77.97 & 655.36 & 638.98 \\
    \end{tabu}
    \vspace{-10pt}
    \label{spec_three}
\end{table}

\begin{figure}[t!]
  \centering
  \includegraphics[width=\linewidth]{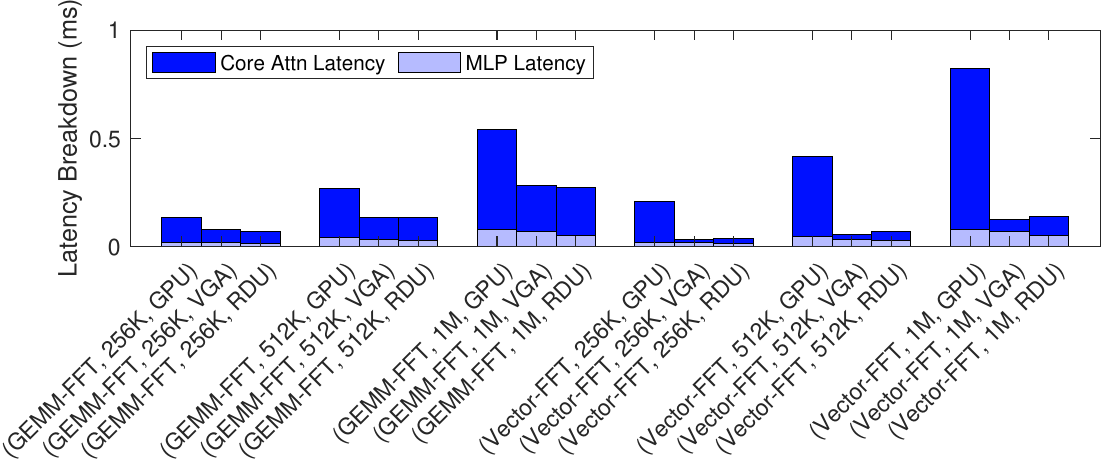}
  \vspace{-10pt}
  \caption{The figure illustrates the latency breakdown of the GEMM-FFT and Vector-FFT variants of the Hyena decoder across three accelerator platforms: GPU, VGA, and RDU.
  For the GEMM-FFT decoder, both VGA and RDU achieve a $2\times$ speedup relative to the GPU, while VGA and RDU achieve similar performance.
  For the Vector-FFT decoder, VGA and RDU provide a $5.95\times$ speedup over the GPU, while VGA and RDU achieve similar performance.}
  \vspace{-10pt}
  \label{hyena_compare}
\end{figure}
We further compare the performance of the RDU architecture against two baselines: an NVIDIA A100 GPU~\cite{nvidiaH100datasheet} and a domain-specific ASIC known as VGA~\cite{lee2024vga}.
The architectural specifications of all three platforms are summarized in Table~\ref{spec_three}.
For a fair comparison, the VGA configuration is scaled to match the compute throughput of the RDU.
In addition, all three architectures are modeled with an 8TB/s HBM3e memory in DFModel~\cite{ko2024dfmodel}.
On the GPU, GEMM-FFT operations are executed on the tensor cores, while Vector-FFT operations are executed on the CUDA cores~\cite{stvrelak2018performance}.
Notably, the tensor cores offer $4\times$ higher compute throughput compared to the CUDA cores, leading to differing performance characteristics depending on the operation type and mapping strategy.
Figure~\ref{hyena_compare} presents the performance comparison results.
For the GEMM-FFT decoder, both RDU and VGA achieve a $2\times$ speedup over the GPU baseline, attributed to their doubled peak GEMM throughput relative to the GPU.
In the case of the Vector-FFT decoder, RDU and VGA outperform the GPU by $5.95\times$, as the GPU has significantly limited throughput available for FFT operations. 
RDU and VGA exhibit comparable performance on the Vector-FFT workload, as both architectures are capable of dedicating their full compute throughput to FFTs.
However, it is important to note that VGA is a fixed-function ASIC designed specifically for FFT and GEMM computations, whereas RDU offers a more general-purpose reconfigurable architecture.
This enables the RDU to support a broader range of workloads that VGA cannot efficiently handle (e.g. Mamba models in Section~\ref{mamba}).

\section{Map Mamba to RDU}
\label{mamba}

This section analyzes scan algorithm variants and their trade-offs.
Based on this, we propose a PCU extension to efficiently support scan operations, and validate its performance through empirical evaluation.

\subsection{Scan Algorithms}

\begin{figure}[t!]
  \centering
  \includegraphics[width=\linewidth]{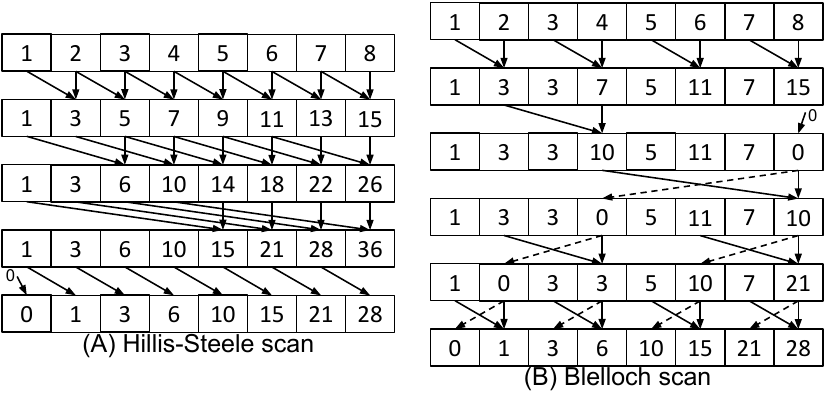}
  \vspace{-10pt}
  \caption{The figure illustrates the dataflow of two parallel-scan algorithms: Hillis-Steele scan (HS-scan) and Blelloch scan (B-scan). HS-scan requires $\log_2{N}$ parallel steps but performs $N \log_2{N}$ total work. In contrast, B-scan completes in $2\log_2{N}$ steps with a total work complexity of $2N$.}
  \vspace{-10pt}
  \label{scan}
\end{figure}

The Mamba decoder relies on an exclusive scan as its core computation kernel~\cite{gumamba, dao2024transformers}.
A commonly used variant is the circular scan (C-scan) algorithm, which appears in domains such as disk scheduling and ultrasonic testing~\cite{worthington1994scheduling, hasiotis2011application}.
The scan computes the prefix sum of a sequence, where each output element is the sum of all preceding input elements. For example, given the input sequence $[2, 4, 6, 8]$, an exclusive scan produces $[0, 2, 6, 12]$.
However, the C-scan algorithm is inherently sequential, computing each output element one at a time.
As a result, it is poorly suited for execution on modern vector accelerators such as GPUs and RDUs.
To address this limitation, parallel-scan algorithms have been proposed to improve hardware efficiency and facilitate mapping to parallel architectures~\cite{Harris2007}.
Specifically, there are two widely used parallel-scan algorithms: Hillis-Steele scan (HS-scan)~\cite{hillis1986data} and Blelloch scan (B-scan)~\cite{blelloch1989scans}, as illustrated in Figure~\ref{scan}.
The HS-scan completes in $log_{2}N$ parallel steps with a total of $Nlog_{2}N$ work.
In each step $i$, every element at index $j$ reads the value at index $j - 2^{i-1}$ (if it exists) and adds it to its own value, propagating partial sums across the array.
This approach has high parallelism but requires more data movement.
The B-scan consists of $2log_{2}N$ steps with a total of $2N$ work, split between an up-sweep (reduction) phase and a down-sweep (distribution) phase.
During the up-sweep, each step computes partial sums in a binary tree fashion, combining pairs of elements to build a tree of cumulative sums.
In the down-sweep phase, the tree is traversed in reverse, and each node distributes the correct prefix value to its children, producing the final exclusive scan result.
% It is worth noting that the Kogge-Stone VLSI fast adder~\cite{kogge1973parallel} shares a similar structure with the Hillis-Steele scan, while the Brent-Kung VLSI adder~\cite{brent1982regular} closely resembles the structure of the Blelloch scan.
% This reflects how the parallel-scan primitive is also widely used in the VLSI domain for designing high-performance adders.
% In general, HS-scan is more step-efficient but less work-efficient, whereas B-scan is more work-efficient but requires more computational steps.
To map a long sequence in the Mamba decoder onto the RDU architecture, we adopt the tiled scan algorithm from~\cite{Harris2007}, which partitions the scan operation into tiles of length $R$, each sized to fit within a single PCU (similar to the FFT tiling approach discussed in Section~\ref{hyena}). 
For end-to-end mapping of the full Mamba decoder in Figure~\ref{ssm} onto the RDU, we employ DFModel~\cite{ko2024dfmodel} to optimally allocate resources across the dataflow graph.
This ensures a balanced on-chip pipeline and maximizes overall throughput.

\begin{figure}[t!]
  \centering
  \includegraphics[width=\linewidth]{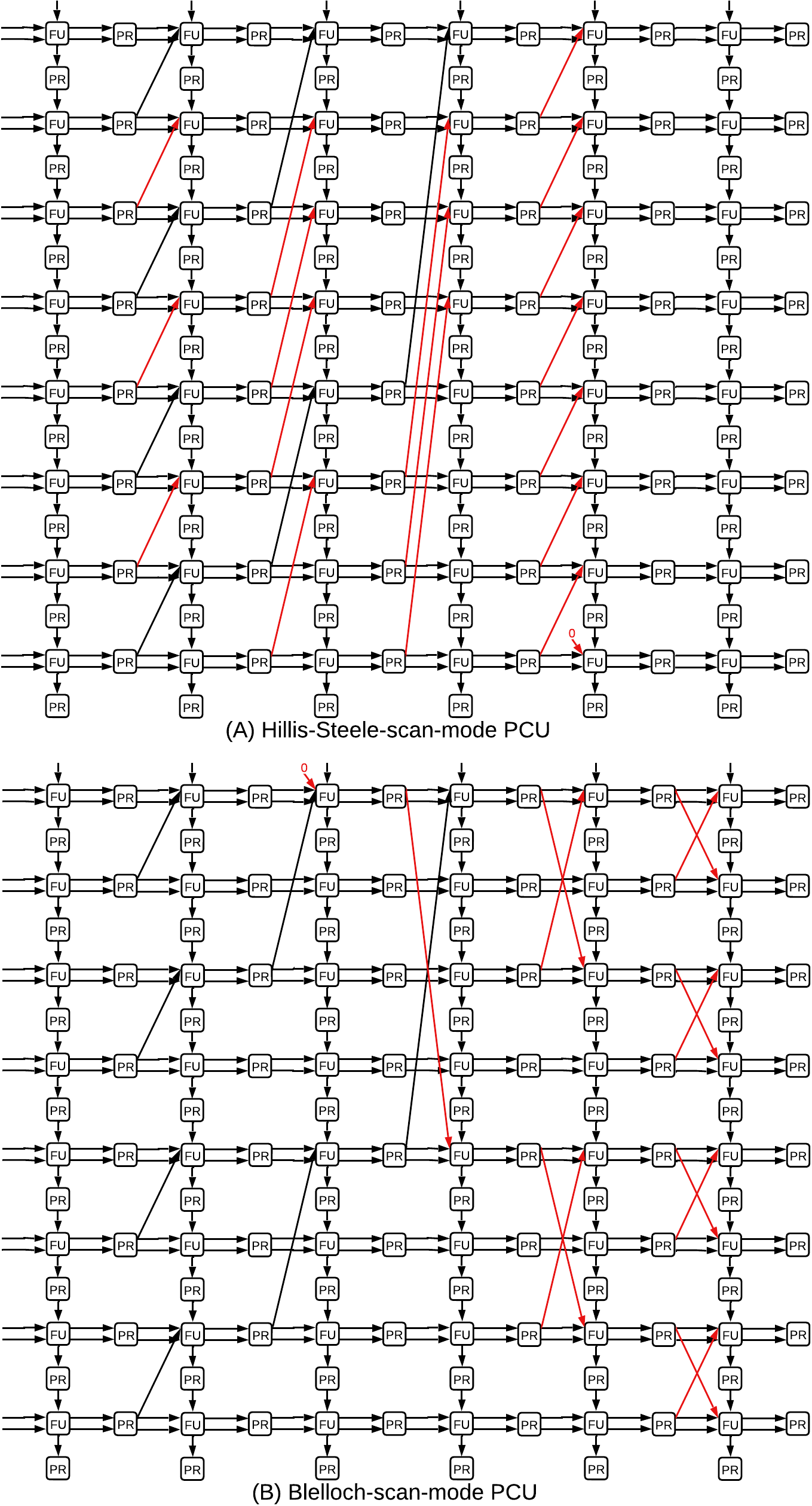}
  \vspace{-10pt}
  \caption{The figure illustrates the Hillis-Steele-scan mode and Blelloch-scan mode of the enhanced PCU.
  The PCU is augmented with cross-lane interconnects between pipeline stages to enable spatial mapping of the parallel-scan algorithms shown in Figure~\ref{scan}.
  This architectural support allows the dataflow of scan operations to be efficiently mapped across the PCU stages, resulting in high performance and utilization.}
  \vspace{-10pt}
  \label{scan_pcu}
\end{figure}

\subsection{PCU Scan Mode}

Executing either HS-scan or B-scan on the baseline PCU architecture shown in Figure~\ref{rdu} is inefficient, as it lacks the necessary cross-lane interconnects required by both parallel-scan algorithms.
To enable efficient execution, we propose an architectural extension illustrated in Figure~\ref{scan_pcu}.
Based on the $8 \times 6$ PCU array, we design two enhanced variants: an HS-scan-mode PCU and a B-scan-mode PCU.
These enhanced PCUs embed the dataflow patterns of the respective algorithms (as shown in Figure~\ref{scan}) directly into the cross-lane interconnect fabric.
This enables spatial mapping of the scan computations onto hardware, significantly improving both performance and efficiency.

\begin{figure}[t!]
  \centering
  \includegraphics[width=\linewidth]{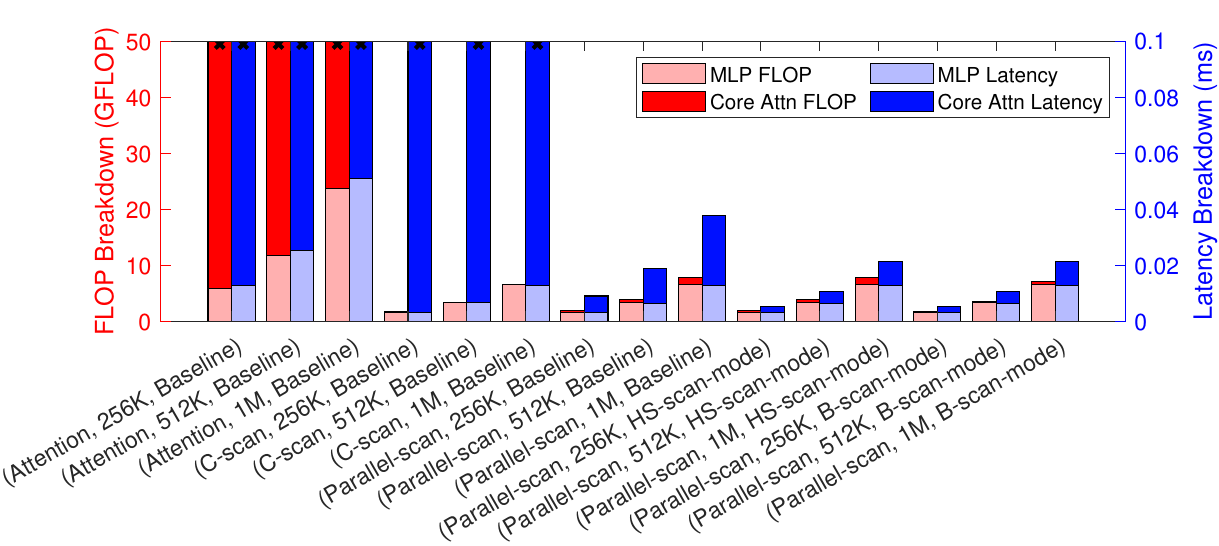}
  \vspace{-10pt}
  \caption{The figure compares five decoder designs: (1) the attention decoder on the baseline RDU, (2) the C-scan Mamba decoder on the baseline RDU, (3) the parallel-scan Mamba decoder on the baseline RDU, (4) the parallel-scan Mamba decoder on the HS-scan-mode RDU, and (5) the parallel-scan Mamba decoder on the B-scan-mode RDU.
  Design (1) exhibits the highest latency due to the quadratic attention computation.
  Design (2) achieves a $7.34\times$ speedup over Design (1).
  Design (3) further improves performance, attaining a $562.98\times$ speedup over Design (2).
  Designs (4) and (5) deliver an additional $1.75\times$ speedup over Design (3).}
  \vspace{-10pt}
  \label{mamba_time_flop}
\end{figure}

\begin{figure}[t!]
  \centering
  \includegraphics[width=\linewidth]{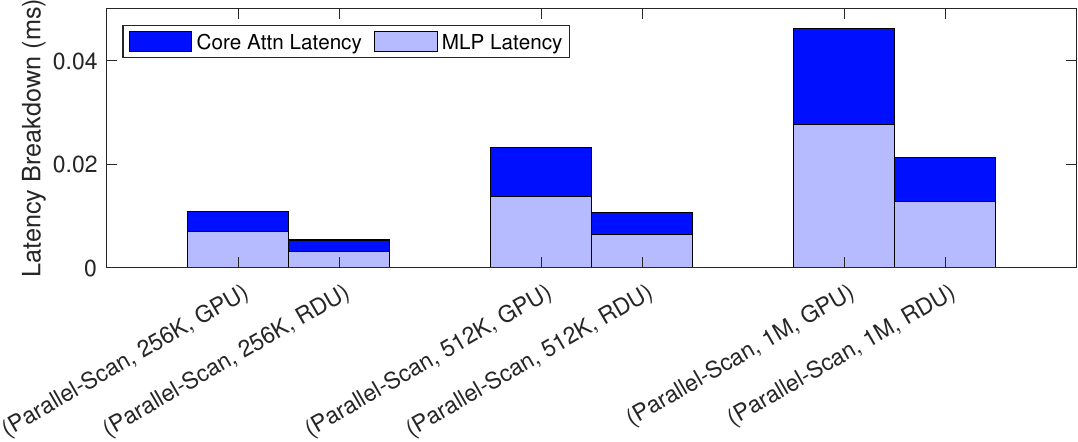}
  \vspace{-10pt}
  \caption{The figure illustrates the latency breakdown of the parallel-scan Mamba decoder on both GPU and RDU.
  The RDU achieves a $2.12\times$ speedup over the GPU.}
  \vspace{-10pt}
  \label{mamba_compare}
\end{figure}

\begin{table}[b!]
    \centering
    \vspace{-10pt}
    \caption{Architectural specifications of two accelerators.}
    \begin{tabu}{p{3cm}p{1.3cm}p{1.3cm}}
    \rowfont{\bfseries}  & GPU & Scan RDU \\
    \midrule
    GEMM FP16 TFLOPS & 311.87 & 638.98 \\
    Scan FP16 TFLOPS & 77.97 & 638.98 \\
    \end{tabu}
    \vspace{-10pt}
    \label{spec_two}
\end{table}

\subsection{Experimental Results}

In our experiments, we evaluate five designs: (1) the attention decoder on the baseline RDU, (2) the C-scan Mamba decoder on the baseline RDU, (3) the parallel-scan Mamba decoder on the baseline RDU, (4) the parallel-scan Mamba decoder on the HS-scan-mode RDU, and (5) the parallel-scan Mamba decoder on the B-scan-mode RDU.
All decoders are configured with a hidden dimension of 32. We sweep across three sequence lengths: 256K, 512K, and 1M. Each design is mapped to an RDU using DFModel~\cite{ko2024dfmodel}, with architectural specifications detailed in Table~\ref{spec}.
 Figure 7
presents the FLOP count and latency breakdown for the five designs.
Several key observations can be drawn from the
figure:
\begin{itemize}
    \item The attention decoder exhibits the highest latency due to the large FLOP count associated with the quadratic attention mechanism.

    \item The C-scan Mamba decoder achieves a $7.34\times$ speedup over the attention decoder, benefiting from the reduced computational complexity of scan operations in Mamba.

    \item The parallel-scan Mamba decoder achieves a $562.98\times$ speedup compared to the C-scan Mamba, as the parallel scan can be effectively vectorized on the RDU architecture, whereas the serial nature of C-scan prevents efficient vectorization.

    \item The parallel-scan Mamba decoder running on either the HS-scan-mode or B-scan-mode RDU achieves an additional $1.75\times$ speedup over the baseline RDU, due to improved pipeline utilization enabled by the enhanced scan-mode interconnects.
    However, the speedup is bounded by the MLP latency, which does not benefit from the scan-mode enhancements, as explained by Amdahl's Law~\cite{amdahl1967validity}.
    Both the HS-scan-mode and B-scan-mode RDUs achieve identical performance, as each mode supports a throughput of one scan per cycle.
\end{itemize}
We further compare the performance of the RDU architecture against an NVIDIA A100 GPU~\cite{nvidiaH100datasheet}, with architectural specifications detailed in Table~\ref{spec_two}. For GEMM operations, the GPU achieves approximately half the throughput of the RDU. For scan operations, the GPU's throughput is only 12\% that of the RDU, as scans are executed on CUDA cores rather than tensor cores~\cite{Harris2007}.
Figure~\ref{mamba_compare} presents the comparison results. The RDU achieves a $2.12\times$ speedup over the GPU, driven by its higher performance in both GEMM operations within the MLP and scan operations in the core attention.

\section{Hardware Overheads}

To evaluate the hardware overheads of the FFT-mode, HS-scan-mode, and B-scan-mode PCUs, we design and implement the three enhanced PCU variants alongside a baseline PCU.
Each PCU is modeled as an $8\times6$ array and described in Chisel~\cite{bachrach2012chisel}. We use \texttt{SInt16} as the data type due to Chisel's limited support for floating-point arithmetic~\cite{stumm2013white}.
The generated Verilog is synthesized using Synopsys Design Compiler~\cite{kurup1997logic} using TSMC 45nm technology~\cite{cheng2007highly} with timing closure achieved at 1.6GHz.
Table~\ref{overheads} reports the area and power overheads of the enhanced PCUs.
In all cases, the overhead is less than 1\% relative to the baseline PCU.

\begin{table}[t!]
    \centering
    \vspace{-10pt}
    \caption{Area and power overheads of the enhanced PCUs.}
    \begin{tabu}{p{2.5cm}p{2.5cm}p{2.5cm}}
    \rowfont{\bfseries}  & Area ($\mu m^2$)  & Power ($mW$) \\
    \midrule
    Baseline PCU &	90899.1 ($1\times$) &	140.7 ($1\times$)\\
    FFT-Mode PCU &	91572.9 ($1.007\times$) &	141.4 ($1.005\times$)\\
    HS-Scan PCU &	91383.0 ($1.005\times$) &	141.2 ($1.004\times$)\\
    B-Scan PCU &	91275.7 ($1.004\times$) &	141.1 ($1.003\times$)\\
    \end{tabu}
    \vspace{-10pt}
    \label{overheads}
\end{table}

\section{Related Work}

% Pimacolaba~\cite{ibrahim2024pimacolaba} leverages processing-in-memory (PIM) architecture to enable efficient FFT computation.
Tang et al~\cite{tang2022high} generate an FFT ASIC using the SPIRAL framework~\cite{franchetti2018spiral}, showcasing a high-throughput design synthesized in advanced technology nodes.
Works such as FlashFFTConv~\cite{fuflashfftconv} and tcFFT~\cite{li2021tcfft} exploit GPU tensor cores to accelerate FFT execution, overcoming limitations of traditional GPU FFT libraries.
Barhen et al.~\cite{barhen2010high} demonstrate an efficient mapping of FFTs onto multi-threaded multicore processors.
Harris et al.~\cite{Harris2007} implement both Hillis-Steele and Blelloch scan algorithms in CUDA, enabling efficient parallel execution on NVIDIA GPUs.
Zegarra et al.~\cite{zegarra2017automatic} propose a framework for the automatic parallelization of scan operations on multicore CPUs using OpenMP.
% Zhang et al.~\cite{zhang2010novel} introduce a novel algorithmic modification to improve scan efficiency on multicore architectures by optimizing data access patterns and reducing synchronization overhead.
However, achieving high efficiency on general-purpose CPUs and GPUs often requires significant algorithmic and software modifications to adapt the computation patterns to the underlying hardware.
Conversely, ASICs can execute these patterns efficiently but require costly and inflexible custom hardware development from scratch.
Our proposed RDU extension bridges this gap by offering both general-purpose programmability and specialized efficiency through lightweight architectural enhancements.

\section{Conclusion}

This paper proposes an architectural extension to the RDU, optimized for efficient execution of long-sequence state-space models including Hyena and Mamba. To overcome the limitations of GEMM-based accelerators such as A100 GPUs and baseline RDUs, we introduce two specialized modes: an FFT-mode for Hyena SSMs and a scan-mode for Mamba SSMs. The FFT-mode achieves a $1.95\times$ speedup over the baseline RDU and $2\times$ to $5.95\times$ speedup over the A100 GPU. The scan-mode achieves a $1.75\times$ speedup over the baseline RDU and a $2.12\times$ speedup over the A100 GPU.
Both extensions incur minimal area and power overheads (less than 1\%), highlighting the RDU's efficiency and architectural flexibility. We hope this work inspires further research into reconfigurable architectures that can support a broader range of computational patterns across diverse workloads.

\bibliographystyle{IEEEtranS}
\bibliography{refs}

\end{document}